# Graphene on Ir(111) characterized by angle-resolved photoemission

Marko Kralj,[1,*] Ivo Pletikosić,[1] Marin Petrović,[1] Petar Pervan,[1] Milorad Milun,[1]

Alpha T. N'Diaye,[2,3] Carsten Busse,[2] Thomas Michely,[2] Jun Fujii,[4] Ivana Vobornik[4]

[1] Institut za fiziku, Bijenička 46, HR-10000 Zagreb, Croatia

[2] II. Physikalisches Institut, Universität zu Köln, Zülpicher Straße 77, D-50937 Köln, Germany

[3] National Center for Electron Microscopy, Lawrence Berkeley National Laboratory, Berkeley, CA 94720, USA

[4] CNR-IOM, TASC Laboratory, AREA Science Park - Basovizza, SS14, km 163.5, I-34149 Trieste, Italy

Angle resolved photoelectron spectroscopy (ARPES) is extensively used to characterize the dependence of the electronic structure of graphene on Ir(111) on the preparation process. ARPES findings reveal that temperature programmed growth alone or in combination with chemical vapor deposition leads to graphene displaying sharp electronic bands. The photoemission intensity of the Dirac cone is monitored as a function of the increasing graphene area. Electronic features of the moiré superstructure present in the system, namely minigaps and replica bands are examined and used as robust features to evaluate graphene uniformity. The overall dispersion of the $\pi$-band is analyzed. Finally, by the variation of photon energy, relative changes of the $\pi$- and $\sigma$-band intensities are demonstrated.

# 1. INTRODUCTION

After the first transport characterization of a single layer of graphite on an insulating support in 2004 [1], graphene has become a material of great scientific and technological interest [2,3,4], especially concerning carbon-based electronics [5]. Its high-mobility and quasi-massless charge carriers originating from the π-, π*-bands in the proximity of the Fermi energy are the foundations of envisaged application of graphene for electronics. Although preparation of graphene by exfoliation enabled enormous fundamental advancement, it is obvious that this method is not suitable to bring graphene to industrial scale sizes. Epitaxial growth, however, can be tuned to large scales. A well established pathway is thermal decomposition of SiC surfaces [4,6,7,8]. The fact that SiC is a wide band gap semiconductor allows direct connection of graphene on SiC wafer to electronics. Another pathway, known for decades, is epitaxial growth on transition metal surfaces by hydrocarbon decomposition in a controlled way [9,10]. In many cases, epitaxial growth enables a production of macroscopic samples [11,12,13,14] and for several systems it has been shown that epitaxial graphene can also be simply removed from its substrate by chemical etching [15,16].

In terms of graphene bonding to a metal, two different scenarios have been proposed, depending on the metal surfaces. Theoretical modeling predicts either a strong chemisorption [e.g. for (111) surfaces of Co, Ni, and Pd] or a weak bonding [e.g. for (111) surfaces of Al, Cu, Ag, Au, and Pt], in which case graphene-metal separations indicate a van der Waals character of the graphene-metal bond [17]. The bonding is intimately related to the Dirac cone, graphene's most relevant intrinsic signature. For the strongly bound systems, the hybridization of the π and π* bands with substrate states causes a bandgap opening in the eV range and corresponding shifts of the bands towards higher binding energy. Such behavior was characterized by angle-resolved photoelectron spectroscopy (ARPES) for e.g. graphene on Ni(111) [18] and Ru(0001) [19]. For weakly bonded systems the Dirac cone is preserved and the electronic structure is close to the one of freestanding graphene. This behavior was found for graphene on Pt(111) [20] and graphene on Ir(111) [21].

On Ir(111) graphene can be prepared with an exceptionally high structural quality [12] and as a single layer. The solubility of carbon in iridium is insignificant at suitable graphene formation temperatures [22]. Thereby graphene growth can be controlled by hydrocarbon supply and self-limited by the availability of uncovered Ir(111) area, which shrinks during growth. This is in contrast to many other transition metals, e.g. ruthenium, which dissolve carbon to a higher extent, where patches of thicker layers are formed almost unavoidably by segregation of carbon from the bulk during cooling [13]. Three preparation methods for graphene on Ir(111) are in use: (i) Low temperature hydrocarbon room temperature adsorption followed by decomposition at a fixed, elevated temperature, referred as a temperature programmed growth (TPG). Flakes perfectly oriented with their dense packed direction in registry with the Ir one are obtained (angular spread 0.25°) [23]. As one TPG cycle yields only a partial coverage around 0.2 monolayers (ML) of Ir(111), larger coverages are only obtained by repetitive TPG cycles [23,21]. (ii) Exposing the hot substrate to hydrocarbon gas is referred to as chemical vapor phase deposition (CVD). CVD enables to cover the entire Ir(111) surface in a single run, but only at very high growth temperatures rotational variants of graphene [24] can be avoided [25]. (iii) A TPG step followed by CVD (TPG+CVD) preserves the excellent orientational order of TPG flakes and simultaneously enables full graphene coverage [12]. With all three methods full to near full coverage with excellent orientation order may be achieved, but the iterative TPG method [method (i)] as well as TPG followed by CVD [method (iii)] are to be preferred if orientational order of graphene is crucial.

The high structural quality makes graphene on Ir(111) appealing for the electron spectroscopy studies. Core level probe, x-ray photoelectron spectroscopy (XPS), was used to monitor the initial stages of graphene growth on iridium. It was proposed that very small graphene flakes have a dome-like shape, with edges strongly bonded to Ir(111) [26]. The fully covered surface was intensely studied by ARPES [21,27,28,29, 30]. The moiré of graphene on Ir(111) is a very good template for the growth of arrays of clusters [31], and part of the ARPES studies exploited the template effect to investigate the superlattice effects in graphene [28,29]. For clean graphene on Ir(111) the measured Dirac cone band structure is largely preserved, but the electronic interaction of graphene with the substrate is reflected through 0.1 eV hole doping, replica bands and a minigap opening in the Dirac cone [21]. Minigaps and replica bands prove the long-range structural quality and uniformity of epitaxial graphene on Ir(111). In that sense ARPES is a very sensitive technique that can be used to compare the quality of differently prepared graphene on Ir(111). All ARPES studies that address clean graphene on Ir(111) largely agree, only with slight discrepancies in the interpretation of spectra near the Fermi energy. Very recently, a synchrotron study with high photon energies of 95 and 130 eV was conducted which enabled to tune the ARPES cross sections for graphene and iridium features near to the K point. Based on this, it was concluded that Ir perturbs the shape of the π-bands very near to the Fermi level, i.e. at the Dirac point, where the Ir 5d surface state, $S_1$ [32], hybridizes with graphene states [30]. This specific electronic interaction was proposed to account for the selection of a single graphene orientation [30]. In addition, a high photon energy of 120 eV used in Ref. [29] suppressed Ir features. Based on these results a Dirac point gap of ≥ 70 meV was suggested.

The present paper has three main topics: First, it complements our previous study [21] by a detailed analysis of the intermediate growth steps leading to a full graphene layer as studied in [21]. Second, it provides data on the electronic structure obtained using the recently refined growth process [12] which leads to well-aligned, nearly defect free graphene on the scale of millimeters. Finally, we will extend our previous study by including also a higher binding energy span for graphene (π- and σ-bands).

## 2. EXPERIMENTAL

The experiments have been performed in three ultra high vacuum (UHV) setups dedicated to ARPES (Zagreb and APE (ELETTRA) beamline [33]) and scanning tunneling microscopy (STM, Cologne). Iridium single crystals of the same purity (99.99%) and orientation accuracy (better than 0.1°) were used. The substrate was cleaned by several cycles of sputtering with 1.5 keV $Ar^+$ or $Xe^+$ ions at room temperature or elevated temperature (1100 K) followed by annealing at 1500-1600 K. Cleanness and quality of Ir(111) were checked by STM (size of terraces, presence of adsorbates or defects) or ARPES (surface states sharpness and intensity [32]) and additionally by low energy electron diffraction (LEED) in all setups. In the TPG experiments described in Sect. 3.A. for each cycle, the surface was exposed to 10 Langmuir (L, 1L=1.33×10$^{-4}$ Pa×s) of ethene to ensure saturation at 300 K and subsequently heated to 1470 K [34]. For the TPG+CVD experiments described in Sect. 3.B. graphene was prepared by one TPG cycle (room temperature ethene exposure 6×10$^{-6}$ Pa for 60 seconds and flash to 1400 K) followed by a CVD run (6×10$^{-6}$ Pa of ethene for 300 seconds while the sample held at 1150 K).

STM images in Cologne were taken at 300 K. Typical tunneling conditions were 1 V bias on the tip and 1 nA tunneling current. STM images are differentiated and appear as if illuminated from the left. In Zagreb, ARPES spectra have been taken by a Scienta SES 100 hemispherical electron analyzer with an overall energy resolution of 25 meV and an angular resolution better than 0.2°. Photons of 21.2 eV from a helium discharge ultraviolet source (beam spot diameter of around 2 mm) were used for the excitation. At APE (ELETTRA), ARPES spectra have been taken by a Scienta SES 2002 analyzer. In the experiment we changed the beam energy in the range 20–80 eV. The typical spot size on the sample was 50×150 μm$^2$. The energy resolution of this setup was in the range of 10–20 meV and the angular resolution 0.2°. During acquisition of ARPES spectra, the sample was cooled to 60 K in Zagreb and to 80 K at ELETTRA. In both setups, the relative azimuth changes were performed by a wobble stick, and were checked according to the orientation of spots in LEED. The base pressure was better than 2×10$^{-8}$ Pa (Zagreb) and 5×10$^{-9}$ Pa [APE (ELETTRA), Cologne].

## 3. RESULTS

### A. Temperature programmed growth

In the following we describe how the morphology and the electronic structure evolve during a series of sequential TPG cycles. Large scale STM images in Fig. 1(a)-(d) show the development of the graphene layer with an increasing number of cycles. Closer inspection and smaller scale images of individual graphene flakes show that they can be found both on the step edges and on terraces, although most of them are attached to the edges. Assuming that the graphene growth rate is proportional to the still uncovered surface area we can make differential Ansatz $d\theta=(1-\theta)\cdot\lambda\cdot dN$. This differential equation is solved and gives the expression for the surface area covered by graphene, $\theta(N)=1-e^{-\lambda N}$, so the graphene coverage exponentially approaches unity with increasing number of growth cycles, $N$. Fig. 1(e) shows the experimentally determined dependence of the graphene coverage on the number of cycles (squares) for one growth series. The data can be fitted by the expression for $\theta(N)$, which gives that $\lambda_{STM}=0.21\pm0.02$.

The flake shapes and sizes vary. From the statistical analysis of STM data we find that the average flake size increases with increasing number of cycles, e.g. after two TPG cycles the typical flake size is $A=9.5\times10^3$ nm$^2$, whereas after seven cycles it reaches $A=7.8\times10^4$ nm$^2$. In order to characterize the resulting morphologies, for each image we also measured the total length of graphene edges divided by the image area, i.e. the graphene step edge density [Fig. 1(e), circles]. This quantity shows a maximum after one cycle due to the large flake density. With increasing coverage, the flake density decreases which clearly indicates that coalescence of flakes is taking place already after two cycles. In the course of TPG growth, wrinkles in the graphene layer are also formed. These wrinkles appear as elongated features in the STM images. In figure 1(c) and (d) several wrinkles are indicated by arrows. They are a sign of relaxation of compressive stress which builds up during cooling due to the different thermal expansion coefficients of Ir and graphene [35], which was also noted for other substrates, e.g. on Pt(111) [20].

Figure 2(a) shows several ARPES spectra focused to a narrow region around the K point of graphene's Brillouin zone (BZ) taken during a preparation which consisted of seventeen TPG cycles. The formation of the Dirac cone centered at 1.7 Å$^{-1}$ can be noted. A thick line (yellow) at the eight-cycle spectrum indicates a position at $k_{\|}=1.784$ Å$^{-1}$ from where the energy distribution curve (EDC) cuts were obtained. Those seventeen cuts are merged to a plot in Fig 2(b) and show the evolution of two clearly resolved peaks: one close to the Fermi energy belonging to the iridium surface state $S_1$ [32] and one at lower energy belonging to the Dirac cone band. Obviously, with increasing number of TPG cycles, the peak intensity characteristic for iridium is reduced while the graphene peak intensity is increased. We also note background intensity at energies below -0.4 eV which gets reduced upon each growth cycle. Making the analysis of peak intensities more quantitative, the EDC curves were best fitted by two Lorentzian peaks of fixed position and width: For iridium at –0.07 eV with a full width at half maximum (FWHM) of 0.13 eV and for graphene at –0.30 eV with FWHM of 0.20 eV. Figure 2(c) shows the areas of the fitted Lorentzian peaks (with linear background subtracted) plotted as a function of the number of TPG cycles. Both the iridium and graphene peak intensities exhibit clear saturation with the increasing number of TPG steps. We assume that the peak area associated with the cut through the Dirac cone, $I_{Gr}$, is proportional to the graphene coverage and following the equation derived for the STM data, the Dirac cone peak area is given by $I_{Gr}(N) \sim (1-e^{-\lambda N})$. For fitting the graphene peak area, the zeroth cycle point was omitted and the value of $\lambda_{Gr}=0.196\pm0.007$ is found. Following the same logic, for the iridium surface state peak we expect a decrease, $I_{Ir}(N) \sim e^{-\lambda N} + I_{off}$, where the offset intensity $I_{off}$ is introduced as the intensity of the surface state must not decay to zero even for a fully covered surface. We find a value of $\lambda_{Ir}=0.45\pm0.01$ with the iridium peak area offset being about 5% of its initial area (at $N$=0). We note that the same $\lambda$ values are obtained if we do not freeze the position of the iridium peak. In that case, the maximum of the iridium peak shifts monotonously with growing number of cycles towards Fermi energy and the overall shift is around 0.025-0.030 eV.

### B. Long-range quality: superperiodicity features and linewidths

In the following we will demonstrate the electronic quality for graphene prepared according to the TPG+CVD process [12]. Figure 3(a) illustrates that graphene prepared by the TPG+CVD method shows very clear superperiodicity signatures: a mini gap (upper arrow) and replica bands, $R_{1-6}$. Figure 3(b) shows a zoom into a minigap region from Fig. 3(a). Another arrow in figure 3(a) points to a position at $k_{\parallel}\approx1.29$ Å$^{-1}$ where obviously another gap opens in the π-band of graphene. This feature is also visible after TPG preparation. This second minigap can be understood in terms of lower lying replica bands approaching the Dirac cone and opening the gap [see Fig 3(c) and discussion].

As already shown in Fig. 2, the ARPES maps can be analyzed by performing energy distribution curve cuts at selected $k_\parallel$ values. Another type of analysis is the momentum distribution curve (MDC) cut at selected energy values. Examples of MDC and EDC cuts are shown in Fig. 4. In each case, the spectral peaks can be fitted by single Lorentzian functions and their full width at half maximum can be determined. Note that the EDC curve in Fig. 4(c) taken at $k_\parallel$=1.46 Å$^{-1}$ represents a cut through the main and replica cone, which illustrates a difference of 6-7 times smaller ARPES intensity of the replica cone band. Systematic EDC or MDC analysis of the Dirac cone can be used to precisely extract the quasiparticle velocity. Positions of peak maxima of EDC and MDC Lorentzians are plotted in Fig 4(d). Their dispersions coincide, except in the narrow minigap region where EDC cuts more clearly resolve the presence of the minigap opening. The dispersions in Fig. 4(d) can be reliably fitted by a linear function in the energy ranges above and below the minigap. The slope of the dispersion between the minigap and Fermi energy is $0.95\times10^6$ m/s. A small deviation from that line near to the Fermi energy can be attributed to the fact that the analyzed spectrum is for the sample azimuth of 0.4°, whereas the K point is at exactly 0° azimuth. Below the minigap, a slightly different value of $1.07\times10^6$ m/s is determined. The small difference in group velocity can be due to the fact that relatively narrow energy regions in the proximity of the minigap are fitted.

### C. Band structure of graphene

By performing scans throughout the Brillouin zone, we map the overall dispersion of the bands for graphene on Ir(111). Figure 5 shows a wide energy scan along the ΓK and ΓM directions of the BZ. The dominant feature in the spectrum is the π-band. The bottom of the measured band at Γ, $E_\Gamma$, is at –8.73 eV and its saddle point at M, $E_M$, is at –2.74 eV. Another feature of the electronic structure of graphene are the σ-bands, which originate from the in-plane hybridized electron orbitals. For a wide range of photon energies the photoemission cross section for these bands is smaller than for the π-band. For a photon energy of 55 eV (figure 5), a dispersion of the σ-band is barely visible. However, as Fig. 6 illustrates, for some photon energies its intensity can be notable. The example in Fig. 6(a) was recorded at a photon energy of 36 eV. The measured σ-band has a saddle point at Γ at around –3.8 eV and exhibits downward dispersion towards K. We observed similarly strong intensity of the σ-band at a photon energy of 70 eV. To illustrate the differences in the ratio of the σ- and π-band intensities, in Fig. 6(b) we compare EDC cuts for two different photon energies at the same $k_{\parallel}$=0.5 Å$^{-1}$ value. Variation of the photon energy also influences the relative intensity of the Ir(111) features, which was already noted for several excitation energies [32].

## 4. DISCUSSION

### A. Temperature programmed growth

From previous STM experiments it is known that each ethene TPG cycle covers a fixed fraction, f, of the bare iridium surface with graphene. The coverage after N cycles can then be expressed recursively: $\theta(N)=\theta(N–1)+f\cdot(1–\theta(N–1))$. This linear recurrence of the first order is solved by $\theta(N)=1–(1–f)^N$ with the starting condition $\theta(0)=0$. This equation can be converted into $\theta(N)=1–e^{-\lambda N}$ using $f=1–e^{-\lambda}$. The constant $\lambda_{STM}$=0.21±0.02 found above then corresponds to a fraction of f=0.19 of free Ir surface covered with graphene after each step. This is in rough agreement with the previous result obtained for one cycle [34]. It has to be noted that the exact determination of the graphene coverage is difficult due to the large size of the resulting structures as compared with the maximum image size accessible for our STM. In the same way, the constant $\lambda_{Gr}$=0.196±0.007 derived from ARPES data gives a value of f=0.178 which is consistent with STM findings and also confirms our assumption that the Dirac cone peak intensity is directly proportional to the graphene covered area. Moreover, the ARPES value is more reliable because it is determined in the measurements with a larger number of successive TPG cycles and the information is derived by integrating over a much larger area of the sample.

The iridium peak attributed to surface state S$_1$ [21,32] follows an exponential drop ~ $e^{-\lambda N}$ where $\lambda_{Ir}$=0.45±0.01 and is suppressed to about 5% of its initial intensity. ARPES data shows that surface states of Ir(111) [32] remain visible at different parts of the BZ, even after formation of a full graphene layer (cf. Fig. 5). This visibility can be understood based on measurements and theory which support the view that graphene on Ir(111) is physisorbed [21,36]. If we assume that the interaction of graphene and the Ir surface state is negligible then we expect that the graphene islands just attenuate the intensity of photoelectrons emitted from the Ir surface state, proportional to graphene coverage. In a simple three-step model picture of the photoemission process, this would reflect the fact that the photoelectron from the iridium surface state has to overcome a larger path and more attenuation before it is emitted from the surface. Very generally, we can account that the inelastic mean free path of the electrons with a kinetic energy of about 15 eV is of the order of 1 nm [37]. When they pass through an additional graphene layer of ~0.34 nm thickness, their intensity is expected to be reduced to about 70%. The measured intensity is only 5%. Therefore we must consider that a more complex mechanism is at work. One mechanism could be the recently proposed specific electronic interaction of the iridium surface state $S_1$ with graphene at the Dirac point [30] which may affect the surface state ARPES intensity at the $k_{//}$ value near to the K point, where we made the EDC cuts. Second, we consider the possibility of an upward shift of the surface state upon graphene adsorption. This scenario is supported by the observation of a small monotonous shift of several tens of meV towards the Fermi level, if the fit position of the iridium peak is not kept frozen for a growing number of TPG cycles. This idea is additionally strengthened by recent scanning tunneling microscopy measurements which have demonstrated that another surface state of iridium, $S_2$ [32], also close to the K point, exhibits an upward shift of about 0.1 eV due to the presence of the graphene overlayer [38]. If a similar effect shifts the $S_1$ state across the Fermi energy, then it is clear that the ARPES intensity we monitor would apparently drop due to the Fermi level cutoff. We also note that much larger surface state shifts have been found and modeled for strongly chemisorbed graphene on Ru(0001) [39]. In this case smaller separation of graphene and the substrate and different symmetry character of the state may be responsible for more pronounced effect.

For all the mechanisms discussed above (attenuation, electronic interaction, upward shift), the intensity of the iridium peak would decay with the increase of graphene coverage as determined by STM, and one would expect the factor $\lambda_{Ir}$ to have a value of around 0.2. The measured value of 0.45, indicates a more rapid drop of the monitored surface state intensity.

We attribute this behavior to scattering of the Ir surface state at graphene edges. Such a scattering was visualized for the surface state $S_2$ [38]. This scattering is enabled by the fact that the edges of graphene flakes are locally more strongly bond to Ir surface [26]. Scattering is therefore related to the relevant structural quantity, namely the step density, which is obtained by STM and displayed in Fig. 1(e). Superimposing a decrease in the intensity of $S_1$ proportional to the graphene covered areas with an additional decrease proportional to the measured graphene step density can qualitatively lead to an overall diminution with a significantly increased $\lambda_{Ir}$ constant, and it is consistent with our observation. In order to get more complete and quantitative description, more STM and ARPES measurements for the TPG growth need to be performed under varying preparation conditions.

**B. Long-range quality: superperiodicity features and linewidths**

The excellent visibility of minigaps and replica bands in figure 3(a) is a direct measure of the structural uniformity of graphene prepared according to the refined growth recipe [12]. In terms of time needed to form the full layer, this method is much faster than TPG growth alone and has some additional advantages: according to STM characterization, the quality of graphene is excellent, with uniform orientation and only few wrinkles. In order to visualize the position and origin of both minigaps seen in experimental data indicated by arrows in Fig. 3(a), a tight-binding approximation (TBA) band structure displaying the Dirac cone and all its six replicas is shown in Fig. 3(c). They are plotted for the azimuth of 0.4°, which fits the experimental data best. The precision of this fit procedure is better than ±0.1° because the Dirac cone and its six replica features are adjusted at the same time by the azimuth. It is obvious that a low energy minigap close to –3eV is due to the same potential that opens the higher energy minigap reported in our earlier study [21].

It is worthwhile to note that the MDC and EDC widths we measure are relatively narrow. In addition to the numbers reported in Fig. 4 we extracted the MDC FWHM at 0.68 eV below the Fermi energy, for comparison to the values presented in Ref. [40]. The measured value of 0.029 Å$^{-1}$ is very similar to the FWHM reported at the same energy for multilayer epitaxial graphene grown on the C face of SiC [40], which is considered as close to isolated graphene. However, the measured widths of epitaxial graphene need to be interpreted with caution. We note that first-principles calculations have been used to simulate the electron interactions and corresponding linewidths of graphene (e.g. Ref. [41]) but the presence of the metallic Ir substrate needs to be taken into account and its relation to the electron correlations in graphene need to be understood. Generally, it is hard to say whether there is an overall quantitative consensus between theory and available experimental data to an extent such as has been achieved for "ordinary" surface states [42]. Finally, our values of the quasiparticle velocity of $0.95\times10^6$ m/s and $1.07\times10^6$ m/s are very close to ARPES slopes reported for epitaxial graphene on other substrates, e.g. Au/Ni(111) [18] or C face of SiC [40].

**C. Band structure of graphene**

The measured positions of the π-band at the Γ and the M point in Fig. 5 cannot be captured by using first nearest-neighbor (1NN) TBA parameters for graphene derived by a fit of density functional theory (DFT) calculations for freestanding graphene [43]. Although the nearest-neighbor hopping energy $\gamma_0$= –2.84 eV and overlap $s_0$=0.07 from Ref. [43] accurately describe the measured π-band dispersion for graphene on Ir(111) very near to the K point [21], due to the known issues in DFT the overall band width is underestimated [44]. By using these parameters we set the bottom of the π-band $E_\Gamma$ at –6.96 eV, and the saddle $E_M$ at –2.56 eV. We have also tested two published sets of 1NN parameters reported to describe the measured π-band for graphene on SiC ($\gamma_0$= –3.28 eV, $s_0$=0.0425) and second layer graphene on Ru ($\gamma_0$= –3.28 eV, $s_0$=0.03) [45,46] and failed to reproduce the observed dispersion of graphene on Ir(111) especially near the Γ and the M point, where the disagreement was larger than 0.15 eV. In order to better describe the whole π-band dispersion within the 1NN TBA, we fit the parameters to best reproduce the measured features. The outcome of this procedure is the dashed line in figure 5, which shows 1NN TBA band fit for $\gamma_0$= –2.848 eV, $s_0$=0.0029, $\varepsilon_{2p}$=0.1 eV.

For a displayed range of binding energies in Figs. 5 and 6, tight binding calculations reproduce two downward dispersing $\sigma_2$ and $\sigma_3$ bands which are degenerate only at the Γ point at about −3.7 eV [47]. Although the relative visibility of the σ feature can be enhanced by appropriate photon energy (cf. Fig. 6), in our spectra we observe only the shallowest σ-band. Its photoemission intensity vanishes at the Γ point but our estimate of −3.8 eV is close to the theory value [47]. This vanishing intensity was already pointed and understood through the role of the symmetry of σ orbitals in the photoexcitation matrix element and also through the cancellation of the amplitudes from two carbon atoms in the graphene unit cell for photoelectron waves at normal emission [47].

## 5. SUMMARY

In summary, the influence of the preparation method of graphene on Ir(111) on its π- and σ-band features have been analyzed by ARPES.

The intensity behavior of the Dirac cone and of the Ir(111) surface state around the K point has been followed during the TPG growth process. The ARPES intensity changes were compared to the STM data for the surface area covered by graphene after a corresponding number of TPG cycles. The Ir(111) surface state ARPES intensity drops significantly for a full layer of graphene, which cannot be explained by intensity attenuation due to increasing graphene coverage only. Possible reasons for this finding are: the specific interaction of the Ir surface state and graphene at the Dirac point [30] and, the upward energy shift of the Ir surface state upon graphene formation [38]. Furthermore, the intensity of the surface state decays faster than the graphene coverage grows. This indicates an additional quenching mechanism which can be found in the scattering from strongly bound graphene edges [26,38]. The increase of the Dirac cone intensity is in agreement with the increase of the graphene area deduced by STM.

Graphene prepared TPG+CVD [12] has been characterized by ARPES. Analysis around the K point of graphene reveals clear signatures of the superperiodic potential, i.e. replica bands and the minigap. In addition to minigaps which were characterized in our earlier study around 1 eV below the Fermi energy [21], in this work we also observe the minigap lower in energy which is due to the same potential. The MDC and EDC widths as well as the measured quasiparticle velocity agree well with the literature for other epitaxial graphene systems.

We characterize the band structure of graphene in the broad energy range along for the ΓK and ΓM directions. The measured dispersion of the π-band can be approximated by a first-nearest neighbor tight-binding approximation. Finally, variation of the photon energy of the p-polarized light reveals that for certain photon energies also the lowest lying σ-band can have relatively pronounced ARPES intensity, comparable to that of the π-band.

## ACKNOWLEDGEMENTS

We gratefully acknowledge financial supports by the DAAD-MZOS via the project "Electronic properties of graphene-cluster hybrids", the UKF by the grant no. 66/10, the MZOS (project no. 035-0352828-2840), the DFG (grant no. Bu 2197/2-1), as well as the support by the CNR-IOM.

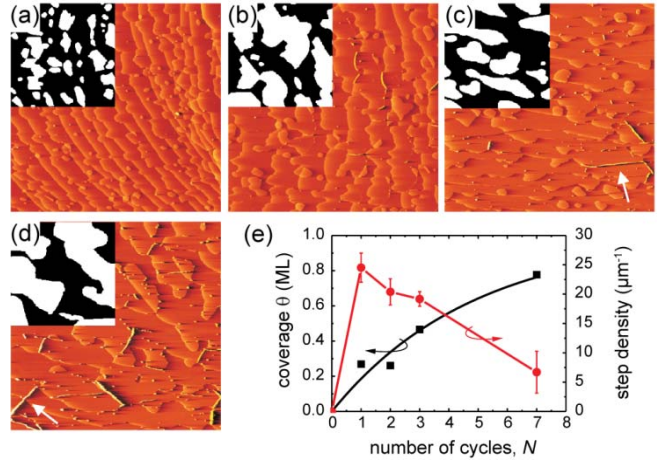

FIG. 1. (Color online) (a)-(d) Set of (1×1) µm² large scale STM images showing the growth of graphene flakes for (a) one, (b) two, (c) three, and (d) seven TPG cycles. In each subfigure, in the upper left quadrant graphene was colored white whereas bare Ir(111) was colored black. White arrows in (c) and (d) point to some of the wrinkles discussed in the text. (e) Squares: surface coverage θ [line is a fit to θ($N$)=1−e$^{-\lambda N}$] as a function of the number of TPG cycles. Circles: graphene step edge density (line to guide the eye), as a function of the number of TPG cycles.

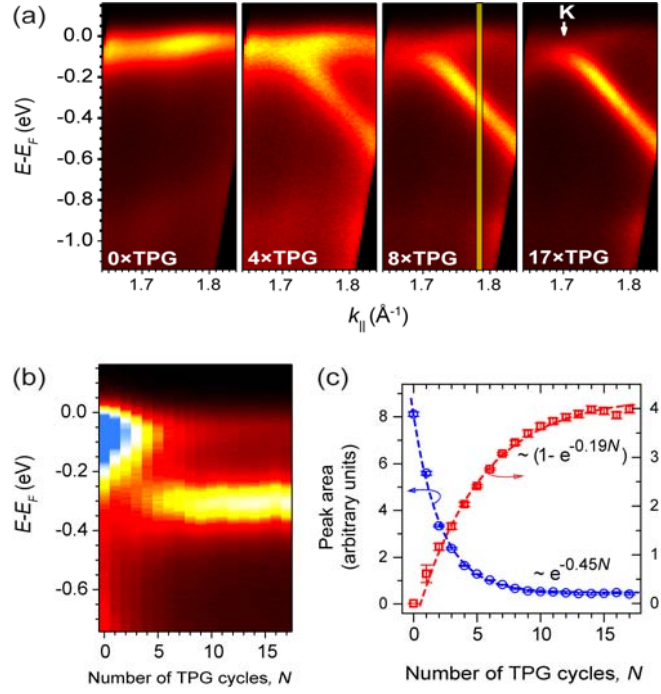

FIG. 2. (Color online) (a) ARPES spectra (excitation energy 21.2 eV, mixed polarization) showing the Dirac cone intensity obtained after the indicated number of TPG cycles. Spectra are focused to the region of the K point of graphene's BZ and were recorded at ~ 60 K. (b) Energy distribution curve cuts obtained from the series of seventeen spectra, part of which are shown in (a), merged into a two-dimensional map. Cuts were taken at $k_{\parallel}$=1.784 Å$^{-1}$ [as indicated by a yellow line in an eight cycle spectrum from (a)]. (c) Photoemission peak areas extracted from spectra in (b). Lines indicate fits of peak areas (see discussion).

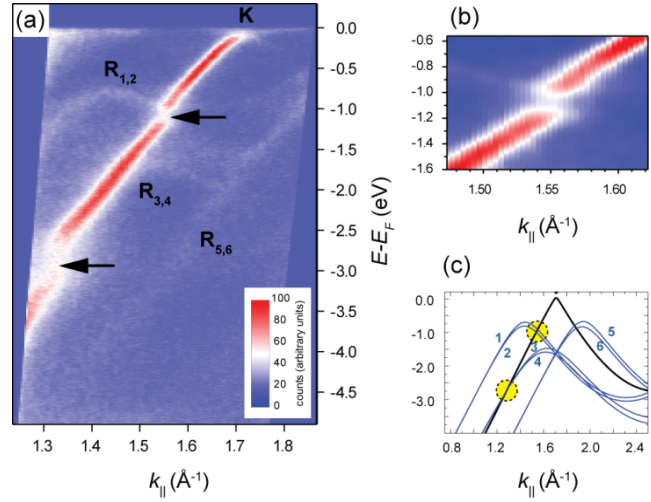

FIG. 3. (Color online) (a) ARPES spectrum (excitation energy 55 eV, p-polarization, sample temperature 80 K) showing the Dirac cone and characteristic features of high quality graphene on Ir(111) discussed in the text. R's indicate replica bands, arrows point to the minigaps and K indicates the position of the K point of graphene. (b) Zoom into the minigap region from (a). (c) Tight binding band of graphene (black) and its superperiodictiy replica bands (blue) for the azimuth of 0.4° that best fit spectrum (a). Numbers 1-6 indicate six replica bands [20]. Circled regions denote regions of minigaps indicated by arrows in (a).

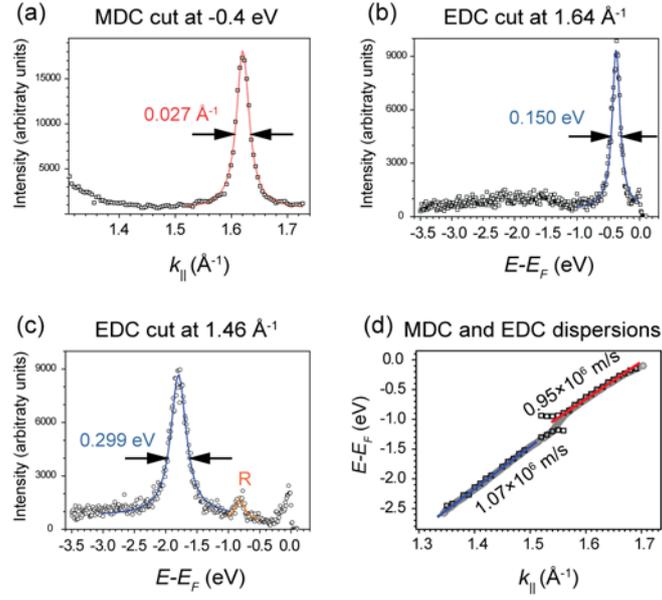

FIG. 4. (Color online) Examples of (a) momentum distribution curve and (b) and (c) energy distribution curve cuts from ARPES spectrum of Fig. 3(a). Positions of cuts as well as measured full widths at half maxima of spectral peaks obtained by Lorentzian fits are indicated. (d) MDC and EDC derived quasiparticle dispersions are shown as full grey circles and open squares, respectively. Lines above (red) and below (blue) the minigap are linear fits of dispersions in the corresponding energy regions and the values of their slopes are indicated.

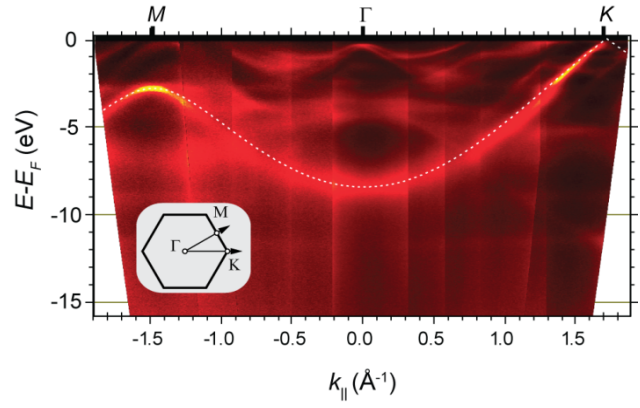

FIG. 5. (Color online) ARPES map showing the dispersion of the π band for the ΓK and ΓM directions. Inset indicates how the ARPES scanning was performed. All spectra shown in this map were acquired at a photon energy of 55 eV (p-polarization) and the sample temperature of 80 K. Dashed line shows the tight-binding calculation fit, discussed in the text.

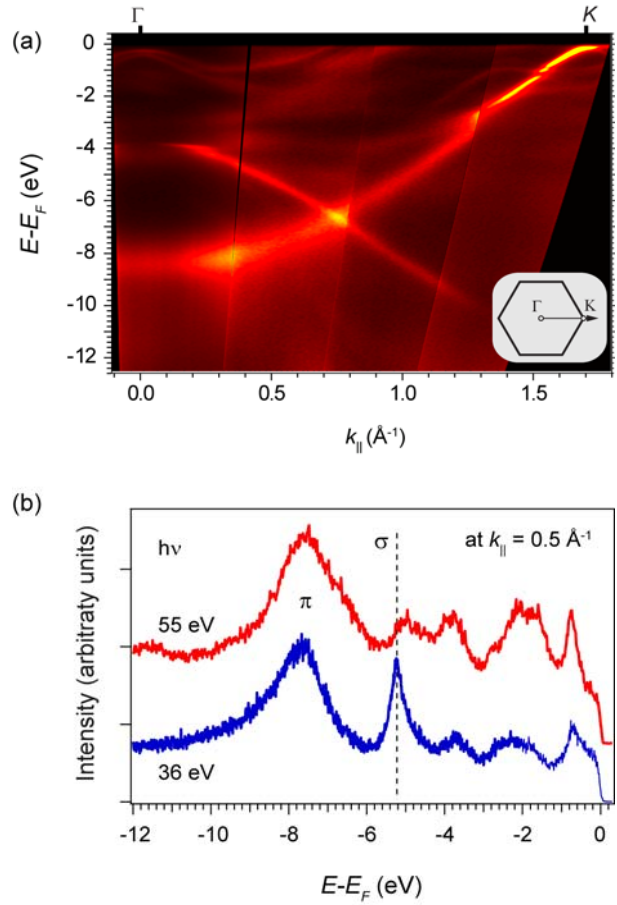

FIG. 6. (Color online) (a) ARPES spectra showing dispersion of graphene bands for the ΓK direction. Spectra shown in this map were acquired at photon energy of 36 eV (p-polarization) and sample temperature of 80 K. (b) EDC cuts at $k_{\parallel}$ =0.5 Å$^{-1}$ for two different photon energies.